\documentclass{emulateapj}
\usepackage{amsmath}
\usepackage{natbib}

\newcommand{\rsun}{{R}_{\odot}}
\newcommand{\msun}{{M}_{\odot}}
\newcommand{\Mmin}{M_{\mathrm{min}}}
\newcommand{\mueff}{\mu_{\mathrm{eff}}}
\newcommand{\Jdot}{\dot{J}}
\newcommand{\Mdot}{\dot{M}}
\newcommand{\Torb}{P_{\mathrm{orb}}}
\newcommand{\Pdot}{\dot{P}_{\mathrm{orb}}}
\newcommand{\Pid}{P_{\mathrm{id}}}
\newcommand{\Kid}{K_{\mathrm{id}}}
\newcommand{\MR}{$M$-$R$ }
\newcommand{\nad}{n_{\mathrm{ad}}}
\newcommand{\nrm}{n_{\mathrm{R}}}

\newcommand{\ee}[2]{\ensuremath{#1 \times 10^{#2}}}

\newcommand{\der}[2]{\ensuremath{\frac{d \,#1}{d#2}}}
\newcommand{\tder}[3]{\ensuremath{\left(\frac{\partial \,#1}{\partial#2}}\right)_{#3}}

\shorttitle{White Dwarf Donors in Ultracompact Binaries}
\shortauthors{Deloye \& Bildsten}

\begin{document}

\title{White Dwarf Donors in Ultracompact Binaries: The Stellar Structure of
Finite Entropy Objects} 
\author{Christopher J. Deloye}
\affil{Department of Physics, Broida Hall, University of California, Santa Barbara, CA 93106}
\email{cjdeloye@physics.ucsb.edu}
\author{Lars Bildsten}
\affil{Kavli Institute for Theoretical Physics and Department of Physics, Kohn Hall,University of California, Santa Barbara, CA 93106}
\email{bildsten@kitp.ucsb.edu}

\begin{abstract}
We discuss the mass-radius ($M$-$R$) relations for low-mass ($M<0.1 \msun$) white dwarfs (WDs) of arbitrary degeneracy and evolved (He, C, O) composition. We do so with both a simple analytical model and models calculated by integration of hydrostatic balance using a modern equation of state valid for fully ionized plasmas.  The $M$-$R$ plane is divided into three regions where either Coulomb physics, degenerate electrons or a classical gas dominate the WD structure.  For a given $M$ and central temperature, $T_c$, the $M$-$R$ relation has two branches differentiated by the model's entropy content.   We present the $M$-$R$ relations for a sequence of constant entropy WDs of arbitrary degeneracy  parameterized by $M$ and $T_c$ for pure He, C, and O. We discuss the applications of these models to the recently discovered accreting millisecond pulsars. We show the relationship between the orbital inclination for these binaries and the donor's composition and $T_c$.  In particular we find from orbital inclination constraints that the probability XTE J1807-294 can accommodate a He donor is approximately 15\%  while for XTE J0929-304, it is approximately 35\%. We argue that if the donors in ultracompact systems evolve adiabatically, there should be 60-160 more systems at orbital periods of 40 min than at orbital periods of 10 min, depending on the donor's composition. Tracks of our mass-radius relations for He, C, and O objects are available through the electronic version of this paper.
\end{abstract}

\section{Introduction \label{sec:intro}}
The discovery of three X-ray transient ultracompact accreting millisecond pulsars (MSPs), XTE J1751-305 \citep{mark02}, XTE J0929-314 \citep{gall02}, and XTE J1807-294 \citep{mark03a, mark03b} have demonstrated the existence of binary pulsar systems with low mass, $M_2 \approx 10^{-2} \msun$, donors.  These three ultracompact systems (here defined as binaries with orbital periods, $\Torb<60$ min) are remarkably homogeneous, with measured $\Torb=42.4,\,43.6,\,40.1$ min respectively, well below the minimum period for a system with a donor composed primarily of hydrogen \citep{rapp82}.  Since the nature of the donors in these systems today depends on the prior evolution of the system, it is useful to discuss the potential formation mechanisms for these systems.

Binary systems with $\Torb<80$ min can form through at least two channels.  Stable mass transfer from an evolved main-sequence star \citep{nels86, fedo89, pods02, nels03} or a He burning star \citep{savo86} onto a neutron star (NS) is one mechanism.  In this scenario, the main-sequence star is brought into Roche lobe contact due to orbital angular momentum losses from magnetic braking at a time when the core has nearly completed H burning.  Such a system will evolve to orbital periods comparable to the ultracompact MSPs and can reach $\Torb \approx 10$ min \citep{pods02, nels03}. \citet{pods02} and \citet{nels03} show that the resulting ultracompact binaries have donor masses  $M_2 \approx 0.1-0.2 \msun$ as they pass through $\Torb\approx 40$ min on their way towards a shorter period.  These masses are significantly greater than those measured in the ultracompact MSPs \citep{gall02,mark02,bild02}.  However, systems evolving through 40 min on the way out from the period minimum have masses more in line with the measurements ($M_2 \approx 0.01 \msun$) and by this time the donors have become partially degenerate with core temperatures $T_c \sim 10^5-10^6$ K \citep{nels03}. 

 The second scenario that may form ultracompact systems involves triggering a common envelope phase during an unstable mass transfer episode from the donor onto the NS. The core of the donor, either a He or C/O white dwarf (WD), and the NS spiral-in to shorter orbital periods until the envelope is expelled \citep{pacz76}. Several authors have proposed binary evolution scenarios in which the system, after emerging from the common envelope phase, then suffers in-spiral due to gravitational wave (GW) emission and eventually re-establishes contact \citep{iben85,rasi00,dewi02,yung02}.  During this long GW in-spiral, the WD will have had time to cool, setting the entropy of the donor at the onset of the second mass transfer phase \citep{bild02}. \citet{taur96} finds that a large fraction of the NS-WD binaries that undergo a common envelope phase will reach contact within 1 Gyr.  Even considering a longer 4 Gyr delay between the formation of the WD secondary and the onset of mass transfer, a He WD will have  $T_c \approx \ee{3}{6}-10^7$ K \citep{alth97,drie99,sere01}, while a C/O WD will have  $T_c \approx \ee{2-3}{6}$ K \citep{sala00}. The mass transfer time-scale at contact is much shorter than the WD cooling time-scale for these WDs so that the  initial entropy of these objects is the minimum attainable.  As noted by \citet{bild02}, if these objects adiabatically expand under mass loss, their $T_c$ will have been reduced by a factor of $\approx 15$ by the time they have reached $M_2 \approx 0.01 \msun$. 

In addition to the evolutionary arguments that donors in the ultracompact MSPs have not reached a $T=0$ configuration, the system XTE J1751-305 provides observational evidence for a hot donor since, as noted by \citet{bild02},  a fully degenerate companion composed of He or C can not fill its Roche Lobe (RL) in this system. Hence, in examining the donors in the ultracompact accreting MSPs, we need to consider them to be arbitrarily degenerate, low mass objects of evolved (He or C/O) composition. To further constrain the nature of these donors (for example, their $T_c$) requires knowledge of their mass-radius, $M$-$R$, relation.  For the compositions (He, C/O), mass ($\sim 0.01 \msun$) and $T_c$ ($\sim 10^5-10^7$ K) ranges of relevance to these objects, the corresponding central densities, $\rho_c$ ($\sim 10^3 \mathrm{g \,cm^{-3}}$) are such that Coulomb and thermal contributions to the equation of state (EOS) provide non-negligible corrections to the degenerate electron pressure, impacting their \MR relations. In this paper, we detail the \MR relation for low-mass stellar objects of finite $T_c$, extending the \MR relations of \citet{zapo69} for $T=0$ objects. In particular, we make clear that there is a \emph{continuous connection} between fully degenerate objects (i.e. WDs), fully convective low-mass stars (i.e. $n=3/2$ polytropes), and Coulomb dominated objects.

 We begin in \S \ref{sec:toymod} by constructing a simple model of these objects using an approximate EOS. Although crude, this model describes adequately the relevant physics and yields an analytic description of the qualitative behavior of the \MR relations and how they are affected by Coulomb and thermal contributions to the EOS. We find that at finite $T_c$, arbitrarily degenerate sequences exhibit a two branch solution, a fact noted previously [e.g.,\citet{cox68, cox64, hans71, rapp84}].  Further, for sufficiently high $T_c$, the sequence of solutions on these two branches exhibit a mutual end point at a \emph{non-zero mass}, $\Mmin$, below which equilibrium solutions do not exist.  When Coulomb contributions are small and electrons are non-relativistic, fully convective stellar models of arbitrary degeneracy are well represented by $n=3/2$ polytropes \citep{haya63, stev91, burr93, usho98}. In \S \ref{sec:polytropes}, we review the role played by degeneracy in determining the \MR relation for $n=3/2$ polytropes and the existence of a two branch solution for the polytrope \MR relation.  Other authors have noted that for a given $M$, there is a maximum $T_c$ that such polytrope models may have \citep{cox68, rapp84, stev91, burr93, usho98}.  We connect the existence of a maximum $T_c$ with that of $\Mmin$.  In \S \ref{sec:adiabats}, we  construct realistic \MR relations using an EOS for fully ionized plasmas. There we exhibit explicitly the impact of Coulomb physics on the \MR relations.  Like the simplified and polytrope models, we find that the \MR relations of this model exhibit a two branch solution and a non-zero $\Mmin$ at high $T_c$. 

In \S \ref{sec:binapps} we apply our stellar models to the ultracompact MSP systems.  For each of these systems, there is a donor (of some composition and $T_c$) which will fill the RL at the required $\Torb$ for any given orbital inclination.  And, for a given composition, there is, in each system, a relation between orbital inclination and $T_c$.  We examine what constraints this places on the composition and $T_c$ of the donors in these systems.  For example, in XTE J1807-294, C and He donors can have any $T_c$, while an O donor's $T_c$ has a minimum value.  In XTE J1751-305, all He or C/O donors must be hot.  We also examine how the future evolution of these systems depends on donor composition and $T_c$ and highlight the fact that a multiple-valued \MR relation leads to a multiple-valued relation between $\Torb$ and the system's mass transfer rate, $\Mdot$.  Finally, in the context of adiabatic evolution, we discuss the expected number distribution of ultracompact systems as a function of $\Torb$.  In a steady state, this distribution depends almost solely on the response of the donor radius, $R_2$, to mass loss through the quantity  $\nrm \equiv d \ln R_2/d \ln M_2$. The increased importance of Coulomb physics in C/O donors alters $\nrm$ as compared to He donors and the expected distribution for the two donor types differs dramatically. Depending on donor type, the relative number of systems at $\Torb \approx 40$ min as compared to those at $\Torb \approx 10$ min is $\approx 60$ for He donors and $\approx 160$ for C/O donors.  We conclude in \S \ref{sec:conc} by discussing future applications of and refinements to our models.

\section{Mass-Radius Relations for Low-Mass Arbitrarily Degenerate Stars \label{sec:models}}
In the mass range of interest ($M_2 < 0.1 \msun$), \MR relations for various objects have previously been constructed. For H rich objects, \citet{burr01} summarize the work that has been done on the structure and evolution of brown dwarfs and related objects.  For these objects, the models include detailed treatment of the equation of state (EOS) and atmospheric physics.  On the other hand, for objects with more evolved composition (i.e. He or C/O) the theory is not so mature.  \cite{zapo69} calculated the \MR relations for $T=0$ objects using the EOS they derived from the Thomas-Fermi-Dirac equation \citep{salp67}.  This EOS treats in an approximate manner the corrections due to Coulomb interactions and exchange effects in a fully-degenerate plasma. Additional \MR relations produced by several different $T=0$ EOS are presented in \citet{lai91}, but do not differ appreciably from the \citet{salp67} results.   For the partially degenerate case, there is a large body of literature for He and C/O core WDs more massive than $0.1 \msun$ \citep{font01, pane00}.  But, only recently have \MR relations for arbitrarily degenerate He and C/O WDs with masses $<0.1 \msun$  been calculated \citep{bild02} and only for a limited number of cases.  Here we fill this gap by constructing low-mass WD models utilizing a realistic EOS. 

\subsection{A Simple Model for Arbitrarily Degenerate Stars \label{sec:toymod}}
Degenerate stellar objects with $10^{-3} \lesssim M/\msun \lesssim 10^{-2}$ have central densities, $\rho_c \sim 10^2$--$10^3 \mathrm{g\, cm^{-3}}$. In this density range, the relative energy contributions to the plasma from an ideal Fermi gas and Coulomb interactions can be comparable. Also, for these densities, at a temperature $T \sim 10^6$ K, the thermal contributions to the pressure are about 10\% of those of the  electrons.  At lower densities, Coulomb and thermal effects become even more significant in calculating the EOS. To examine the interplay between these contributions and how they impact the structure of low-mass stars, we start with a simple EOS for a plasma composed of ions with charge number $Z$ and atomic mass number $A$,
\begin{equation}
P (\rho,T) = P_e (\rho) + \Pid(\rho, T) + P_C(\rho)\,,
\label{eq:Psimp}
\end{equation}
where $P_e$ is the pressure of a fully degenerate non-interacting Fermi gas of electrons $P_e = K_e \rho^{5/3}$, $\Pid$ is the pressure of an ideal gas of ions and electrons, $\Pid = \Kid \rho T$, and $P_C$ is the negative pressure contribution due to Coulomb interactions in the Wigner-Seitz approximation, $P_C = - K_C \rho^{4/3}$. Here $K_e =\ee{3.323}{12} (2/\mu_e)^{5/3}\,\mathrm{dyne\, cm^3\, g^{-5/3}}$ ($\mu_e$ is the mean molecular weight per electron and equals $A/Z$ in a single composition plasma), $\Kid = \ee{8.25}{7} (1+Z)/A\; \mathrm{dyne\, cm\, g^{-1} K^{-1}}$, and $K_C = \ee{2.23}{12} Z^{2/3} (2/\mu_e)^{4/3} \mathrm{dyne\, cm^2\, g^{-4/3}}$.

Consider a one-zone stellar model, i.e. a spherical system characterized by a single pressure and density, $P$ and $\rho$.  From dimensional analysis, $P \sim G M^2/R^4$ and $\rho \sim M/R^3$ where $M$ and $R$ are the mass and radius of the star and $G$ is the Newtonian constant of gravitation.  With the pressure given by equation (\ref{eq:Psimp}),  
\begin{equation}
 G\frac{M^2}{R^4} \approx \Kid T \frac{M}{R^3} +K_e\frac{M^{5/3}}{R^5}-K_C \frac{M^{4/3}}{R^4}\,.
\label{eq:mrsmast}
\end{equation}
In the $T \rightarrow 0$ limit, the term involving $\Kid$ vanishes and solving for $R$ gives
\begin{equation}
 R(T=0) =\frac{K_e M^{1/3}}{K_C +G M^{2/3}}\,,
\label{eq:mrst0}
\end{equation}
showing that the gravitational and Coulomb attraction act to collapse the star, which is supported by the degenerate electron pressure.  As $M \rightarrow 0$, Coulomb forces dominate gravity and $R \propto M^{1/3}$.  As $M\rightarrow \infty$, Coulomb forces become negligible and $R \propto M^{-1/3}$. The transition from gravitational to Coulomb dominated regime occurs where $M \sim (K_C/G)^{3/2} \sim 10^{-4} Z (2/\mu_e)^2 \msun$.  The relation $R \propto M^{1/3}$ in the low $M$ limit implies a constant density, $\rho_{\mathrm{min}}$, fixed by the balance between the electron pressure and Coulomb attraction. Exhibiting $\rho$ explicitly as a function of $M$, $\rho = (G M^{2/3} + K_C)/K_e$ and $\rho_{\mathrm{min}} = (K_C/K_e)^3 = 0.671 Z^2\;\mathrm{g\, cm^{-3}}$.

For $T>0$, equation (\ref{eq:mrsmast}) has solutions given by
\begin{equation}
\begin{split}
R&= \frac{M^{1/3}}{2 \Kid T} \biggl(K_C +G M^{2/3}\pm\\&\sqrt{\left(K_C^2 -4 K_e \Kid T\right)+2 K_C G M^{2/3}+G^2 M^{4/3}}\,\biggr)\;.
\label{eq:mrsimp}
\end{split}
\end{equation}
\begin{figure}
\plotone{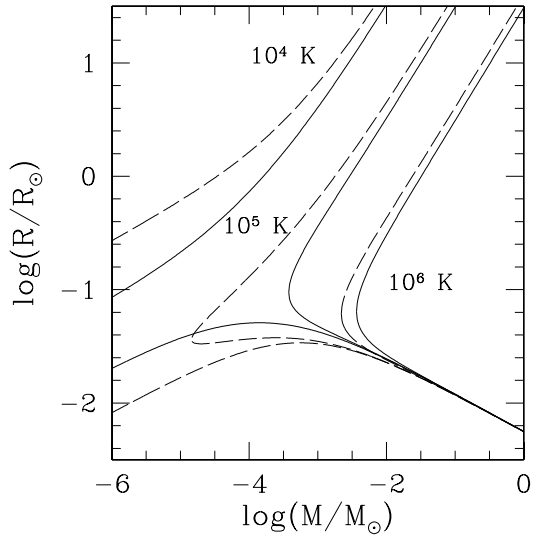}
\caption{The schematic \MR relation of our simple model given by equation (\ref{eq:mrsimp}).  The curves show isotherms at $T=10^4,\, 10^5,$ and $10^6$ K for pure He (solid) and C (dashed) compositions. The curves on the upper branch are labeled with their respective temperatures.  The lower branch curves at low $M$ have a $T=10^4$ K.   \label{fig:mrsimp}}
\end{figure}
Figure \ref{fig:mrsimp} exhibits several isotherms of the \MR relation of equation (\ref{eq:mrsimp}).  The two-branch nature of this relation is apparent, as is the clear separation into three classes of stellar objects.  The large $M$  lower-branch is made up of gravitationally bound objects supported by degeneracy pressure.  On the lower-branch at small $M$, we have Coulomb dominated objects. The upper branch consists of objects supported by thermal pressure.

For $T>0$, there exists a minimum mass, $\Mmin$, found by setting the discriminant in equation (\ref{eq:mrsimp}) to zero,
\begin{equation}
\Mmin= \frac{2 (K_e \Kid T)^{1/2}-K_C}{G}\,.
\end{equation}
This expression is only positive if $T > \ee{4.53}{3} A Z^{4/3}/(1+Z) (2/\mu_e)$ K.  For He(C) this temperature is $\ee{1.5}{4}\, (\ee{8.5}{4})$ K.  Above this critical temperature, the two branches meet at a mutual end point and no solution with $M<\Mmin$ exists. When Coulomb physics is negligible, $M_{\mathrm{min}}$ occurs at the point where $P_e=\Pid$ along the solution curve. 
The existence of $\Mmin$ results from the fact that for $M < \Mmin$, the pressure provided by $P_e+\Pid$ at any $\rho$ is greater than that needed to support the star (this is not the case for either an ideal gas or a Fermi gas independent of the other---in either of these cases it is well known that equilibrium solutions down to $M=0$ exist).  Alternately, for a given $M$, the isotherm on which $M=M_{\mathrm{min}}$ gives the maximum $T$ for which solutions exist with this mass.  For an object starting out on an upper branch solution, e.g. a recently expired star, as it loses entropy via radiation, it contracts.  For a fixed mass, a star supported by thermal pressure has $T \propto R^{-1}$, and it heats up, as expected. The thermal pressure in this case goes as $\Pid \propto R^{-4}$, but $P_e \propto R^{-5}$ and eventually $P_e$ dominates, halting substantial contraction.  Further entropy loss leads to a reduction in $T$.  For a given $T$ and $M$, the two possible solutions are physically distinguished by their entropy.

\subsection{Polytrope Models for Low-Mass Stars Neglecting Coulomb Physics \label{sec:polytropes}}
Ignoring non-ideal effects and assuming that the electrons are non-relativistic, a fully convective object has an EOS obeying $P \propto \rho^{5/3}$ and is modeled by an $n=3/2$ polytrope.  The specific entropy, $s$, throughout such a model is a constant. As $s$ only depends on the degeneracy parameter $\eta$, defined as the ratio of the chemical potential of the electrons to $k T$ ($k$ being the Boltzmann constant), $\eta$ is also a constant of the model \citep{usho98}.  Following \citet{usho98}, this allows us to write the pressure of a non-interacting gas of electrons and ions as
\begin{equation}
P = \frac{\rho}{\mueff m_p} k T\,,\label{eq:polyP}
\end{equation}    
where $m_p$ is the mass of a proton and $\mueff$ is defined as
\begin{equation}
\frac{1}{\mueff} = \frac{1}{\mu_i}+ \frac{2 F_{3/2}(\eta)}{3 \mu_e F_{1/2}(\eta)}\,,
\end{equation}
thus varying as $\eta$ changes. Here $\mu_i$ is the ion mean molecular weight and $F_k$ is the Fermi-Dirac function of order $k$ from \citet{cox68}.  We then utilize the $n=3/2$ polytrope relations, equation (\ref{eq:polyP}), and the fact that $F_{1/2}(\eta) \propto \rho/(\mu_e T^{3/2})$ to determine $M$ and $R$ as a function of $\eta$, $T_c$.  These results can be expressed as
\begin{align}
&\frac{R}{\rsun} = 0.359 \left(\frac{M}{0.01 \msun}\right)^{-1/3} \left[\mu_e^{2/3} \mu_{\mathrm{eff}} F_{1/2}^{2/3}\right]^{-1}\label{eq:polyR}\,,\\
&T_c = \ee{3.42}{5}\, \mathrm{K} \left(\frac{M}{0.01 \msun}\right)^{4/3} \left[\mu_e^{1/3} \mu_{\mathrm{eff}} F^{1/3}_{1/2}\right]^2\label{eq:polyT}\,,
\end{align}
which is a minor rewrite of the results in \citet{usho98}.  

Equation (\ref{eq:polyT}) shows the relation between $T_c$ and $M$ is a function of $\eta$ through the combination $\mu_{\mathrm{eff}} F^{1/3}_{1/2}$, which has a single maximum at $\eta \approx 3-5$ for expected WD interior compositions.  Just as in the simple models of \S \ref{sec:toymod}, this shows explicitly the connection between a maximum $T_c$ for a given $M$ and the existence of an $\Mmin$ for a fixed $T_c$.  Here again, there are 2 equilibrium radii for a given $M$ and $T_c$ differentiated by their degree of degeneracy or equivalently their entropy.

The transition from a thermal pressure dominated to degeneracy pressure dominated state in these models approximately determines $\Mmin$. This transition occurs near where $1/\mu_i \approx$ $2 F_{3/2}/3 \mu_e F_{1/2}$. For increasing $\mu_i$, this occurs at lower $\eta$, i.e. at lower $\rho_c$ if $\mu_e$ and $T_c$ are held fixed.  With $P \propto \rho^{5/3}$, dimensional analysis shows $M \propto \rho_c^{1/2}$; a lower density at the transition between degenerate and non-degenerate states gives a smaller $\Mmin$.  For a fixed $T_c$, a pure C WD will has a smaller $\Mmin$ than a pure He WD.

\subsection{Mass-Radius Relations for Isentropic White Dwarfs \label{sec:adiabats}}
We now calculate the \MR relations derived from stellar models that include Coulomb physics.  In these calculations, we assume that the interior profile of our stellar models are adiabatic.  In reality, the actual entropy profile in a given donor will depend on its evolutionary history.  Many factors---such as whether the system formed through a stable or unstable mass transfer channel, whether or not H burning is still ongoing at the point of contact, and how the mass transfer rate changes with time--- can impact either the initial entropy profile of the donor or its subsequent evolution.  In calculating models for donors in ultracompact binaries, without choosing their evolutionary history, a reasonable approximation of their internal entropy profile is the best that can be done.  Since we aim to construct models that will enable analysis of the donor's properties today, irrespective of their past histories, we must assume some internal profile in calculating them.

A system that initiates mass transfer at a $\Torb \lesssim 40$ min has a donor mass $M_2 \gtrsim 0.01 \msun$.  The mass transfer time-scale for such a system with a NS primary is roughly $M_2/\Mdot \lesssim 1-100$ Myr, depending on $M_2$.  Consideration of the flux through the half-mass point in our models due to electron conduction (calculated using the conductive opacities of \citet{pote99}) compared to the heat content of the interior half of the model shows roughly that the time required to transport this heat out of the interior is $\sim 100$ Myr-1 Gyr, again depending on $M_2$.  Thus during the mass transfer episode that would lead up to the creation of systems at $\Torb \approx 40$ min, the internal evolution is to first order an adiabatic expansion and, in the absence of tidal heating, the initial entropy profile should be more or less preserved with some corrections due to heat transport. To what ever degree this occurs, the critical point is that the interior will not be able to maintain an isothermal profile. We chose to use an adiabatic profile, instead of another arbitrary choice, since it provides several convenient features---it is a limiting case for the possible thermal profiles of the donor and produces the most compact configuration for a given $T_c$ and $M_2$, it is completely determined by the utilized EOS, and it allows parameterization of a set of models by one quantity, the specific entropy which is constant throughout the model.  In addition, the calculations of \citet{nels03} show that He donors tend to become adiabatic as they lose mass.  These latter calculations also highlight that throughout the mass loss episode, donors remain far from isothermal except in the deep interior of the star, and then only at masses near 0.1 $\msun$ \citep{nels03}.

\subsubsection{Equation of State \label{sec:ad:eos}}
\begin{figure}
\plotone{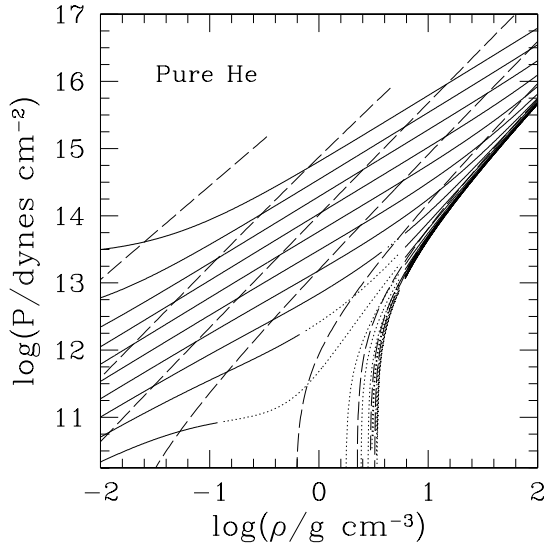}
\caption{The He EOS we utilize showing the comparison isotherms (solid/dotted lines) and a set of representative adiabats (dashed lines). The isotherms are for temperatures incremented by $\Delta \log(T/\mathrm{K}) =0.5$, with the upper curves corresponding to $\log (T/K)=7$. Along the isotherms, the dotted lines indicate regions where the plasma is not fully ionized. The adiabats typically cross into regions where full ionization of the plasma is not definite and is a source of uncertainty in some of our models. \label{fig:eos_ad}}
\end{figure}

For the equation of state, we use the results of \citet{chab98} and \citet{pote00}.  Their work provides accurate prescriptions for calculating the EOS for a fully ionized plasma in either liquid or solid states and which includes the ideal contributions from non-degenerate ions, degenerate electrons and the non-ideal contributions due to Coulomb interactions.  We also include the radiation contribution, a small effect.  In the calculations below, we assume that the plasma is in the solid state when the quantity
\begin{equation}
\Gamma \equiv \frac{(Z e)^2}{a k T} \geqslant 173\,,
\end{equation}
 \citep{faro93}.  Here $a$ is the inter-ion spacing given by $a = (4 \pi n_i/3)^{-1/3}$, $n_i$ being the number density of ions. The quantity $\Gamma$ is a measure of the strength of the ionic Coulomb interactions.  We display a set of isotherms for a pure He plasma calculated with this EOS in Figure \ref{fig:eos_ad} with the solid and dotted lines.  The dotted lines indicate regions in parameter space where $T$ and $\rho$ are such that the plasma will likely become only partially ionized.  In these regimes this EOS is not strictly valid.  

The dashed lines in Figure \ref{fig:eos_ad} show representative adiabats for this EOS, some of which cross into regions of partial ionization. Since we utilize an adiabatic internal profile in calculating our models in \S \ref{sec:ad:calc}, in some of these models there can be a point in their outer layers where our EOS becomes invalid.  We use our EOS in these regimes despite this problem for two reasons.  First, in the models of most interest only the very most outer layers of the models lie in regimes where ionization state transitions become an issue and the calculated \MR relation is not affected.  Second, this EOS provides a simple method of calculating the EOS for various compositions over a wide range of $\rho$ and $T$ and this ease of use would be sacrificed by constructing composition specific extensions to the EOS.  We will highlight in our results models in which these EOS concerns may cause more than a few percent uncertainty in the calculated \MR relations.

\subsubsection{Calculation of the Models and Results \label{sec:ad:calc}}
We constructed models for arbitrarily degenerate objects by integrating mass conservation and hydrostatic balance while presuming an adiabatic temperature profile.  The calculation of our models proceeds as follows.  In the interior, where degeneracy pressure dominates, we integrate  
\begin{equation}
\der{T}{r} = \frac{\Gamma_2-1}{\Gamma_2} \frac{T}{P} \der{P}{r}\,,
\end{equation}
where $\Gamma_2$ is the adiabatic exponent
\begin{equation}
\frac{\Gamma_2}{\Gamma_2-1} = \tder{\ln P}{\ln T}{\mathrm{ad}}\,,
\end{equation}
found from the EOS. The density at each integration step is then solved for from $P$ and $T$ using the EOS.  At low $\rho$, where Coulomb effects on the pressure become significant, we switch to integrating
\begin{equation}
\der{\rho}{r} = \frac{1}{\Gamma_1}  \frac{\rho}{P} \der{P}{r}\,,
\end{equation}
where 
\begin{equation}
\Gamma_1 =  \tder{\ln P}{\ln \rho}{\mathrm{ad}}\,,
\end{equation}
is another adiabatic exponent, and solve for $T$ at each integration step from $P$ and $\rho$.   Each model integration was terminated once the following criteria were met---(1) between two integration steps in which $P$ differed by more than a factor of $\approx 20\%$, $m$ and $r$ changed by no more than 1 part in $10^{8}$, and (2) the current pressure is such that $P/P_c < 10^{-8}$.

We change the integrated variable because as $P\rightarrow0$ along an adiabat, $\rho$ becomes very insensitive to $P$ and it is numerically intractable to determine $\rho$ by root finding from $P$ and $T$.  On the other hand, in the low $P$ limit, determining $T$ accurately from $P$ and $\rho$ is possible, something that was not true in the highly degenerate regime.  We switch from integrating $T$ to integrating $\rho$ when a rough measure of the degeneracy, $167\,\rho/\mu_e T^{3/2} < 100$, is satisfied (where $\rho$ and $T$ are in cgs units).  In the solid state, when $\Gamma \gg 180$, $\Gamma_2 \gg 1$ due to the rapid decline in the plasma's specific heat once crystalline.  We use an adaptive step-size explicit Runge-Kutta algorithm to integrate our equations with the step size chosen to maintain a fixed fractional accuracy. A dramatic increase in $\Gamma_2$ causes a sharp decline in the speed of the integration as the algorithm tries to maintain this accuracy in all three integrated quantities, $m$, $P$, and $T$.  To deal with this problem, once $T =100$ K, we set $dT/dr \approx 0$.  This causes no issues in the \MR relations because by this point we are already well in the $T\rightarrow0$ limit as far as the $P$-$\rho$ relation is concerned in any of our models.

Finally in these models the ion coupling parameter, $\Gamma$, increases with $r$.  This is  due to the fact that along an adiabat, $\Gamma_3-1=(d\ln T/d\ln \rho)_{ad}$ varies from $2/3$ to $\approx 1/2$ in a Coulomb plasma.  Since $\Gamma \propto \rho^{1/3}/T$, and $T \propto \rho^{\Gamma_3-1}$, $\Gamma$ goes as $\rho$ to a negative power.  This is not just true for adiabatic profiles---any object with a profile $T \propto \rho^\gamma$ for $\gamma > 1/3$ will have $\Gamma$ increasing with $r$ (assuming $Z$ is fixed)---because of this some models transition from liquid to solid in their outer layers.  For these cases we do not attempt to match adiabats in the solid and liquid phase (i.e. we do not account for the latent heat).  Instead, the integration assumes continuity of $P$ and $T$; the entropy in these models have a small discontinuity.  If crystallization of the object were to actually occur from the surface inward, this could have significant impact on the mass loss rate, since the primary's gravitational field would have to overcome the Coulomb binding of the crystal to effect mass transfer.  More realistic calculations are obviously needed to consider this further and consideration of this potential effect in evolutionary calculations are encouraged.

\begin{figure}
\plotone{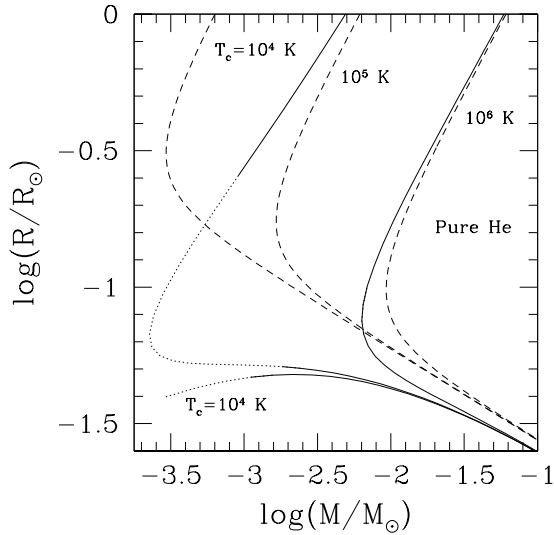}
\caption{A Comparison between \MR relations for polytropes and the full adiabatic models with Coulomb physics for pure He.  The short dashed curves show the \MR relation for $n=3/2$ polytropes of pure He at $T_c= 10^4, 10^5, $ and $10^6$ K.  The solid curves show He WDs calculated with the full EOS of \S \ref{sec:ad:eos} at the same set of $T_c$ as the polytropes.  The significance of Coulomb interactions at the masses shown on the \MR relation is obvious. The dotted portions of the full curves indicate models where more than 5\% of the model's mass is located in regions where the EOS is not strictly valid.   \label{fig:polycomp}}
\end{figure}

\begin{figure}
\plotone{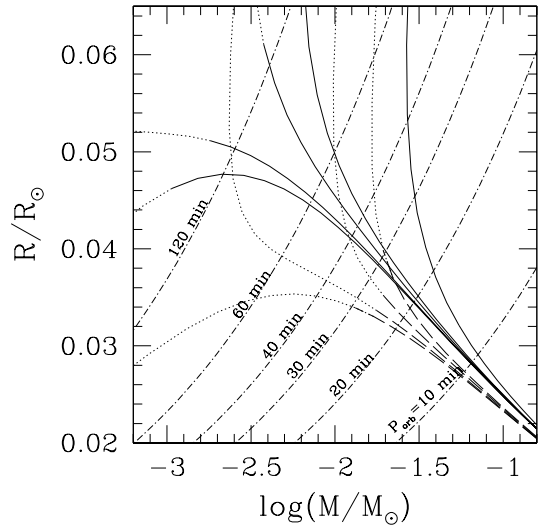}
\caption{The \MR relations for adiabatic models composed of He (solid lines) and C (dash lines). The dotted portions of these curves indicate models where more than 5\% of the model by mass is is located in regions where the EOS is not strictly valid.  The He curves show models with $T_c =10^2,\,10^5,\,\ee{5}{5},\,10^6$ and $\ee{5}{6}$ K. The C curves, models with $T_c =10^4,\,10^6,\,\ee{3}{6},\,$ and $\ee{5}{6}$ K.  The dot-dash lines are the $M-R$ relations for objects filling their Roche lobes at the noted orbital period.  \label{fig:headmr}}
\end{figure}

Typical results for our model calculations are shown in Figures \ref{fig:polycomp} and \ref{fig:headmr}. Figure \ref{fig:polycomp} shows for pure He models how the results utilizing the EOS of \S \ref{sec:ad:eos} differ from the polytropes that neglect Coulomb physics.  In this, and in Figure \ref{fig:headmr}, portions of the \MR curves shown in dotted lines indicate models where more than 5\% of the mass lies in regions where our EOS is not strictly valid. The impact of Coulomb interactions on the structure of low mass WDs is clear from the comparison of the He polytrope models (short dashed lines) to the realistic EOS He models (solid lines).   
Figure \ref{fig:headmr} displays a set of \MR isotherms for our He (solid lines) and C (dash lines) models along with lines of constant $\Torb$ for a donor filling its Roche lobe overlaid (dash-dot lines with $\Torb$ indicated).  Again, the dotted portions of the \MR relations indicate models in which more than 5\% of the model's mass lies in a regime where the EOS is not strictly valid.  From Figures \ref{fig:polycomp} and \ref{fig:headmr}, it is clear that the realistic treatment of Coulomb physics in the EOS is necessary to calculate accurately the structure of low mass WDs. 

\section{Application of the Models to Ultracompact Accreting Millisecond Pulsars \label{sec:binapps}}
\begin{figure}
\plotone{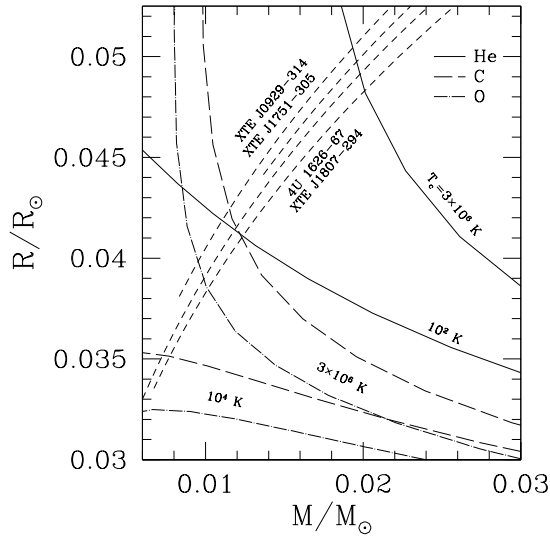}
\caption{A comparison between the \MR relations for our adiabatic He (solid lines), C (long dashed lines), and O (dash-dot lines) along with \MR relations for Roche lobe filling donors in the three known accreting millisecond pulsars and 4U 1626-67 (short dashed lines).  The He \MR relations have $T_c=10^2$ and $\ee{3}{6}$ K; The C and O, $T_c=10^4$ and $\ee{3}{6}$ K. The low $T_c$ relations show approximately the $T=0$ relation for each composition. The RL filling solution for 4U 1626-67 extends down to $M=0$ to indicate that this system has not yet had its mass function measured.\label{fig:amspRL}}
\end{figure}

\begin{figure}
\plotone{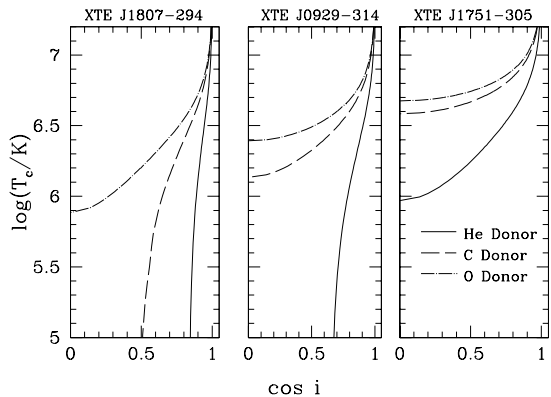}
\caption{The relation between the orbital inclination, $i$, and $T_c$ of our adiabatic He, C, and O donors in the ultracompact systems.   Hot donors are required in XTE J1751-305 for either He or C/O WD donors; C/O donors in XTE J0929-314 must be hot, while for XTE J1807-294 a $T=0$ C/O donor is allowed if the system is nearly edge on. \label{fig:mtsini}}
\end{figure}

We now apply the adiabatic models of \S \ref{sec:adiabats} to the three known ultracompact accreting millisecond pulsars and the high-field X-ray pulsar 4U 1626-67.  In Figure \ref{fig:amspRL} we display (short dashed lines) the \MR relations of Roche lobe (RL) filling donors in XTE J0929-314, XTE J1751-305, and XTE J1807-294 \citep{mark02, gall02, mark03a, mark03b} and \MR relations for our adiabatic models.  We show two isotherms each for He, C, and O models, an approximate $T=0$ \MR relation and one for hot ($T_c=\ee{3}{6}$ K) models.  If the donors in all three systems are He WDs, then $T=0$ objects are allowed in XTE J0929-314 and XTE J1807-294, but XTE J1751-305 requires a hot donor \citep{bild02}.  For C/O donors, whose \MR relations will lie between the C and O models shown, only XTE J1807-294 permits a RL filling cold donor.  The other two systems both require hot C/O donors.  The curves showing the RL filling \MR relations are parameterized by the orbital inclination, $i$ (where $i=0$ is a face on system) and there is a correspondence between $T_c$ and $i$ for each donor composition, as shown in Figure \ref{fig:mtsini}.  From Figure \ref{fig:mtsini}, the donor in XTE J1751-305 must have $T_c \gtrsim 10^6$ K. From orbital inclination constraints, the probability that XTE J1807-294 can accommodate a  He donor is 15\% (for XTE J0929-304, it is $\approx$ 35\%). 

 The other ultracompact accreting pulsar, 4U 1626-67, at $\Torb=41.4\ {\rm min}$ \citep{midd81,chak98} is also shown in Figure \ref{fig:amspRL}. Though the orbit has not yet
been detected by timing the pulsar, the current upper limit of
$a_x\sin i < 8$ lt-ms \citep{levi88, chak97} allows us, in conjunction with our
theoretical work, to constrain the nature of the donor star.  Ever
since the discovery of a neon emission line \citep{angel95}
from this system, there have been active discussions of the nature of
the donor. The \citet{schul01} measurement of a high Ne to O ratio
(further inferred in other ultracompacts by \citet{juett01}), led them to suggest that
the donors in these binaries are the cores of previously crystallized
C/O WDs. \citet{home02} have since also seen strong C and O
lines, but no evidence for helium. Hence, this system seems a likely
one to use for probes of C/O donors. 

For any $\sin i$, a star that fills the RL at the measured $\Torb$ for each of the ultracompacts can always be found by some combination of entropy, composition, and  mass and the current values of each impact the orbital evolution of the system. For $M_2/M_1 < 0.8$, the Roche radius, $R_L$, can be approximated by  \citep{pacz67}
\begin{equation}
R_L \approx 0.46 a \left(\frac{M_2}{M_1+M_2}\right)^{1/3}\,,
\label{eq:RL}
\end{equation}
where $a$ is the separation between $M_1$ and $M_2$.  Combined with Kepler's third law, this leads to the so-called period-mean density relationship, 
\begin{equation}
\Torb \simeq 8.9\,\mathrm{hr} \left(\frac{R_2}{\rsun}\right)^{3/2} \left(\frac{\msun}{M_2}\right)^{1/2}\,.
\label{eq:pdrel}
\end{equation}
Assuming conservative mass transfer, the mass transfer rate (a \emph{positive} quantity) is given by \citep{verb93}
\begin{equation}
\frac{\Mdot}{M_2} =2 \frac{\Jdot}{J} \frac{1}{\nrm+5/3 - 2 M_2/M_1}\,,
\label{eq:mdot}
\end{equation}
where $J$ is the orbital angular momentum, $\Jdot$ the angular momentum loss rate set by GW emission \citep{land62} and
\begin{equation}
\nrm \equiv \der{\ln R_2}{\ln M_2} \,.
\label{eq:n}
\end{equation}
denotes how the donor's radius changes with mass loss.

\begin{figure}
\plotone{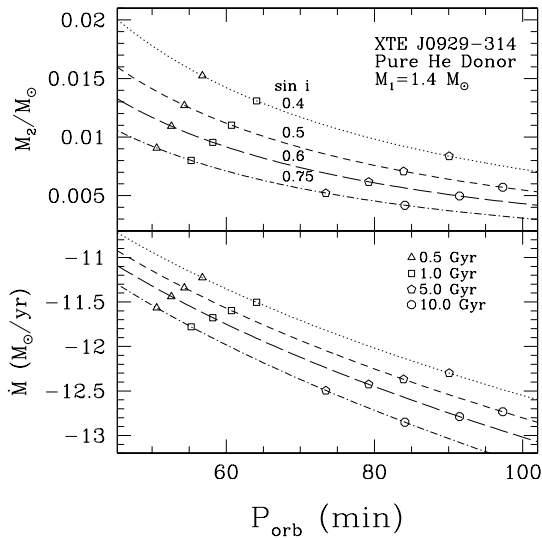}
\caption{The relation between $M_2$, $\Mdot$ and $\Torb$ for XTE J0929-314 over the course of its orbital evolution forward from today as it depends on $\sin i$, or equivalently, the entropy of the donor.  In particular, the $\Mdot-\Torb$ relation is parameterized by the donor's entropy.  Along each curve, the symbols designate time from today: triangles, 500 Myr; squares, 1 Gyr; pentagons, 5 Gyr; circles, 10 Gyr. A smaller $\sin i$ requires a more massive donor and will produce a faster evolution in $\Torb$ for the system. \label{fig:hej0929T}}
\end{figure}

For a given system, $\Jdot$ will depend on the inclination through $M_2$ and $a$.  The rate at which the orbit evolves will vary accordingly as will the $\Mdot-\Torb$ relation over the course of the evolution.  To illustrate this, we calculated the forward evolution of XTE J0929-304 assuming a He donor and four different $\sin i$ values.  We assumed that the NS has $M_1=1.4 \msun$ (and ignore the change in $M_1$ due to accretion) and set 
\begin{equation}
\nrm=\nad \equiv \left(\der{\ln R_2}{\ln M_2}\right)_{\mathrm{ad}} \,,
\end{equation}
so that the donor evolves adiabatically, ignoring any heating mechanisms (such as irradiation or tidal heating) and cooling.  We show the results in Figure \ref{fig:hej0929T}, displaying $M_2$ and $\Mdot$ as a function of $\Torb$.  These relations are not single valued, but parameterized by orbital inclination, or equivalently, by the donor's entropy.  A smaller $\sin i$ requires a more massive donor (which must be hotter than a lower mass donor if it is to have the fixed mean density implied by the system's $\Torb$; see also Figure \ref{fig:mtsini}).  This gives a higher $\Mdot$ for a fixed $\Torb$, as seen in Figure \ref{fig:hej0929T}.  This also impacts the rate at which the orbit evolves.  In Figure \ref{fig:hej0929T}, the age of the system from today is indicated along each curve by symbols and it can be seen that the smaller $\sin i$ the faster the system will evolve in $\Torb$.

\begin{figure}
\plotone{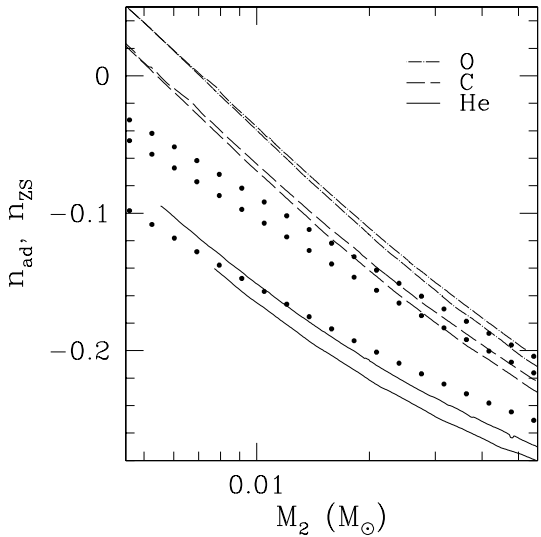}
\caption{The adiabatic change in donor radius with respect to mass, $\nad$ for He (solid lines), C (dashed lines), and O (dashed-dotted lines) stars along two representative adiabats for each composition.  The lower curve in each set has $T_c=10^7$ K at the high mass end; the upper curve has $T_c=10^2$ K (i.e. it is effectively the $T=0$ sequence for each composition). The stronger Coulomb physics decreases the magnitude of the radius response to mass loss in the C/O WDs as compared to He WDs.  The large dots show $\nrm$ for the $T=0$ Zapolsky-Salpeter models for He, C, and O (bottom to top).  \label{fig:nad}}
\end{figure}

\begin{figure}
\plotone{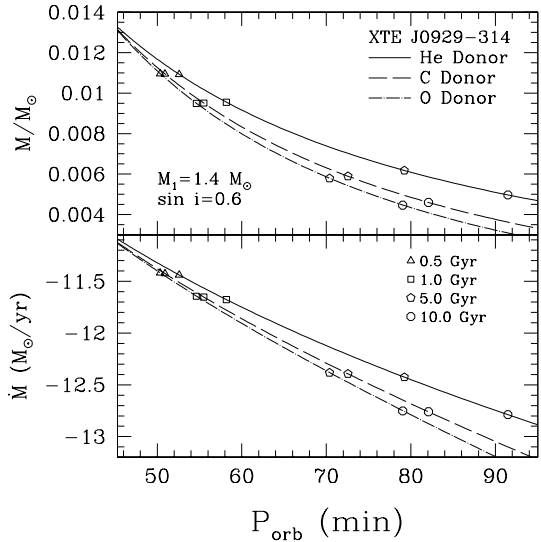}
\caption{A comparison between the evolution of XTE J0929-304 assuming different donor types for an orbital inclination of $\sin i=0.6$.  Shown are $M_2$ and $\Mdot$ as a function of $\Torb$ for He, C, and O donors.  The contrast in the initial evolution between the three donor types comes from the differences in their $\nad$.  The symbols mark the age of the system from today.  \label{fig:aj0929T}}
\end{figure}

At a given orbital inclination in a specified system, a C/O donor must have a higher $T_c$ to fills the Roche lobe than a He donor.  This is due to the stronger Coulomb physics in the C/O object, which also causes $\nad$ to differ between the donor types and impacts the binary's evolution.  The difference in $\nad$ between composition is shown in Figure \ref{fig:nad} for two representative adiabatic tracks for each composition.  The solid dots show the Zapolsky-Salpeter $\nrm$ for the same compositions. The effect of different $\nad$ on orbital evolution is evident in Figure \ref{fig:aj0929T}, which compares the evolution of XTE J0929-304 for $\sin i=0.6$ and He, C, and O donors. The difference in the tracks comes about due to the difference in $\nad$ due to composition; higher $Z$ donors evolve fastest in mass, but slowest in $\Torb$ because they remain more dense than lower $Z$ donors at a given mass.

\section{The Period Distribution for Adiabatically Evolving Ultracompact Systems \label{sec:numdist}} 
These evolution calculations highlight the dependence of observables ($\Torb$ and $\Mdot$) on the donor's entropy and composition.  We now emphasize the impact of the donor composition on the ultracompact population, especially the resulting orbital period distribution. In the scenario of cooling WDs reaching contact via in-spiral (e.g. \citet{nele01}), the relative number
of He vs C/O WD's that come into contact and stably reach longer orbital periods is hard to know. However, what we will show here is the ability to constrain the relative populations of say, He donors, at one orbital period versus another. 

\begin{figure}
\plotone{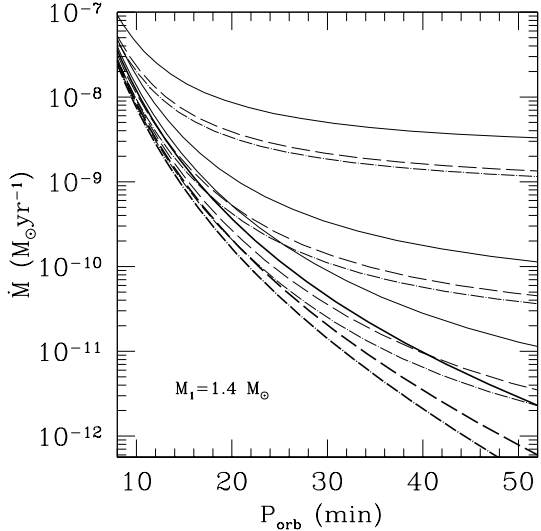}
\caption{The $\Mdot$-$\Torb$ relations assuming $\nrm=\nad$ for He (solid lines), C (dash lines), and O (dash-dot lines).  For each composition, the four curves show lines of constant $T_c =10^2,\,\ee{3}{6},\, 10^7$, and $\ee{3}{7}$ K (bottom to top).  The bolder lines are the $T=10^2$ K curves. \label{fig:ns_ucsys}}
\end{figure}

 A more complete picture of this dependence is shown in Figure \ref{fig:ns_ucsys},  where we display for He (solid lines), C (dashed lines), and O (dash-dot lines) donors, the $\Mdot$-$\Torb$ relation along \MR isotherms with $T_c=10^2,\,\ee{3}{6},\,10^7$, and $\ee{3}{7}$ K, assuming $M_1=1.4 \,\msun$ and $n=\nad$.  These are instantaneous $\Mdot$ along an adiabatic track at a given $\Torb$ and $T_c$.  One can see immediately that for a given $\Torb$, $\Mdot$ can constrain the combination of donor $T_c$ and composition.  In particular, a sufficiently strong upper limit on $\Mdot$ can \emph{rule out} a He donor for a given $\Torb$.  Above the minimum $\Mdot$ for a He donor, further information about the donor composition is difficult to infer without constraints on $T_c$.

\begin{figure}
\plotone{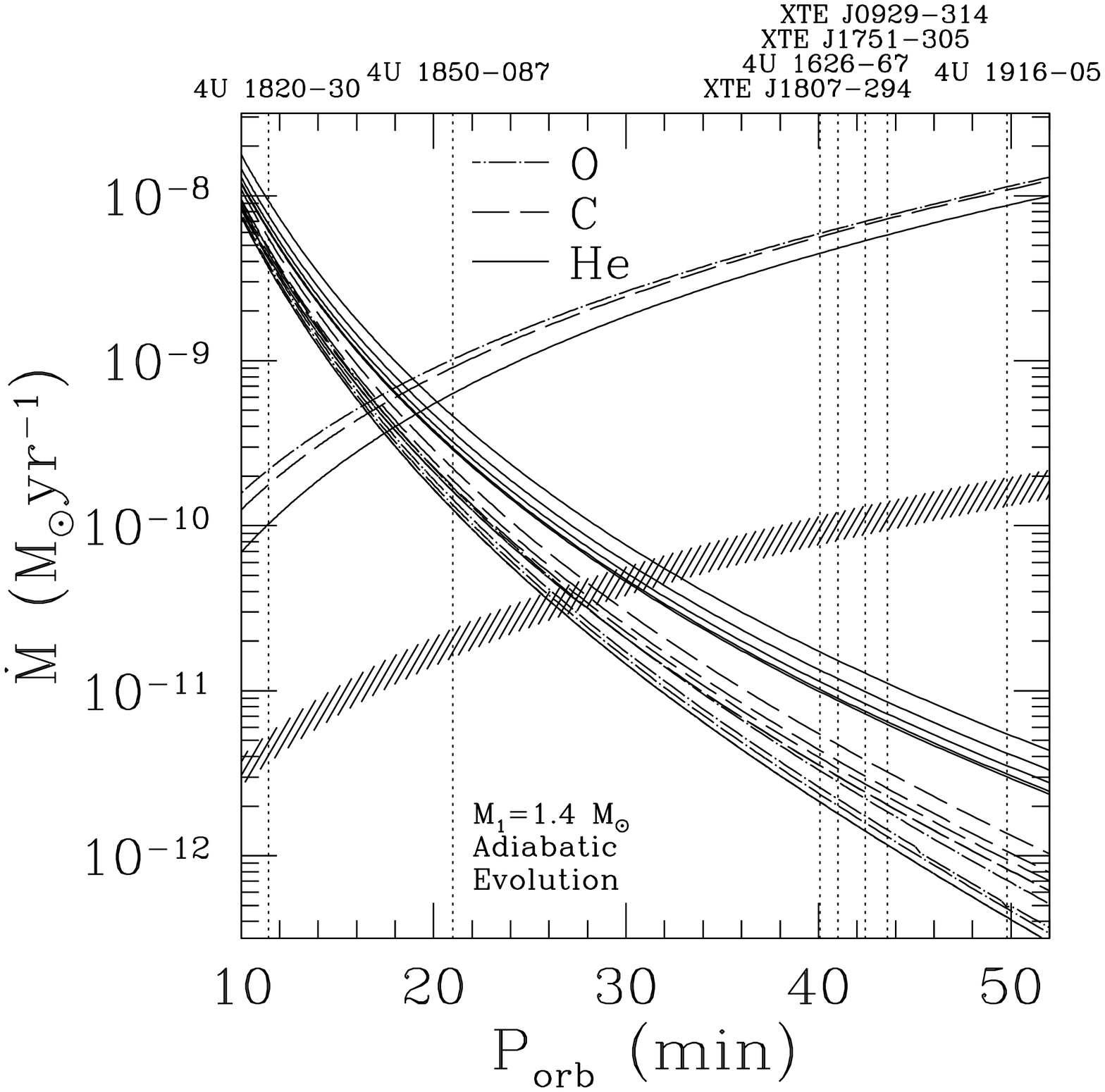}
\caption{The $\Mdot$-$\Torb$ relations along adiabatic evolutionary tracks. Each track starts with a donor filling its Roche lobe at $\Torb=10$ min.  For each composition---He (solid lines), C (dash lines), and O (dash-dot lines)---tracks for models with initial $T_c = 10^2,\,\ee{3}{5},\,\ee{2}{6},\,\ee{5}{6}$ and $10^7$ K are shown (bottom to top). By the time these systems have evolved to $\Torb > 30$ min,  $\Mdot \lesssim 10^{-10} \msun\,\mathrm{yr^{-1}}$; donors in binaries with $\Torb \gtrsim 30$ min that have persistent $\Mdot$'s higher than this cannot have adiabatically evolved from $\Torb \leq 10$ min.  The vertical dotted lines show the orbital periods for ultracompact systems with a NS primary.  For each composition, the upward sloping lines show the critical $\Mdot$ below which the accretion disk is thermally unstable ignoring irradiation \citep{meno02} . The  shaded horizontal band gives the critical $\Mdot$ for an irradiated disk \citep{dubu99} for a range of irradiation efficiencies. The band corresponds to the range of values for the \citet{dubu99} $C$ parameter of $\pm 50 \%$ the fiducial value. \label{fig:adiabmdotp}}
\end{figure}

Now consider adiabatic evolution  with initial $M_1=1.4\, \msun$ and donors of varying composition and $T_c$ which fill their RL at $\Torb=10$ min.  We evolve these systems assuming the donor responds adiabatically to mass loss.  The resulting tracks in the $\Mdot$-$\Torb$ diagram are shown in Figure \ref{fig:adiabmdotp}, along with the measured periods of the known ultracompact binaries with a NS primary (vertical dotted lines) and the critical $\Mdot$ below which the accretion disk in these systems is subject to thermal instabilities for both irradiated (hatched region, \citet{dubu99}) and non-irradiated disks (nearly horizontal lines, \citet{meno02}).  As compared with Figure \ref{fig:ns_ucsys}, it can be seen that for systems at $\Torb>30$ min to have \emph{time-averaged} mass transfer rates $\Mdot > 10^{-10} \msun$ yr$^{-1}$, the donor cannot have evolved adiabatically from systems coming into contact at $\Torb \approx 10$ min. There are potentially two examples of such systems: 4U 1626-67 ($\Torb=41.4$ min and $\Mdot>\ee{2}{-10}\,\msun\,\mathrm{yr}^{-1}$ \citep{chak97}) and 4U 1916-05 ($\Torb=49.8$ min, $\Mdot\approx\ee{5}{-10}\,\msun\,\mathrm{yr}^{-1}$ \citep{swank84}). If these measured $\Mdot$s reflect the long-term average $\Mdot$, then the donors must be extremely hot ($> 10^7$ K, see Figure \ref{fig:ns_ucsys}). However, the typical $\Mdot$ at these orbital periods are below where a He or C/O disk is thermally stable (see Figure \ref{fig:adiabmdotp}; \citet{meno02, dubu99}).  In that case we would explain the present  $\Mdot$ as a higher rate indicative of a system in outburst.  Indeed, the luminosity from 4U 1626-67 is observed to be in a steady slow decline \citep{mavr94,chak97}.

If the ultracompacts evolve adiabatically, we can determine their relative numbers as a function of $\Torb$.   Defining $N(\Torb)$ such that $N\,d\Torb$ is the number of systems with orbital period between $\Torb$ and $\Torb+d\Torb$ and demanding continuity gives
\begin{equation}
\der{(N \Pdot)}{\Torb} = 0\,,
\label{eq:cont}
\end{equation}
leading to the expected relation between $\Pdot$ and $N$:
\begin{equation}
\frac{N}{N_0} = \frac{\Pdot,_0}{\Pdot}\,,
\label{eq:NofPdot}
\end{equation}
where $N_0$ and $\Pdot,_0$ are the respective quantities at some reference orbital period, $\Torb,_0$. For $\nad$ fixed, $R \propto M^{\nad}$ and with $M_2 \ll M_1$, equations (\ref{eq:RL}), (\ref{eq:pdrel}), and (\ref{eq:NofPdot}) lead to the simple relation
\begin{equation}
\der{\ln N}{\ln \Torb} = \der{\ln \Pdot}{\ln \Torb} = \alpha \equiv \frac{11/3-5 \nad}{1-3 \nad} \,.
\label{eq:dlnNdlnP}
\end{equation}
In this case
\begin{equation}
\frac{N}{N_0} = \left(\frac{\Torb}{\Torb,_0}\right)^\alpha\,,
\label{eq:powlaw}
\end{equation}
and from this it is clear that \emph{$\nad$ alone} determines the number distribution.  As $\alpha$ increases with $\nad$, systems with C/O donors will have a stronger increase in $N$ with $\Torb$ than those with He donors due to the difference in $\nad$ shown in Figure \ref{fig:nad}.  In the more general case, $\nad$ is variable and equation (\ref{eq:dlnNdlnP}) becomes, up to terms of order $M_1/M_2$,
\begin{equation}
\der{\ln N}{\ln \Torb} =\frac{11/3 -5 \nad}{1-3 \nad} +\frac{18}{(3 \nad + 5)(3 \nad-1)}\der{\nad}{\ln \Torb},
\label{eq:gen_dlnNdlnP}
\end{equation}
and in general the distribution is almost solely a function of $\nad$.  

\begin{figure}
\plotone{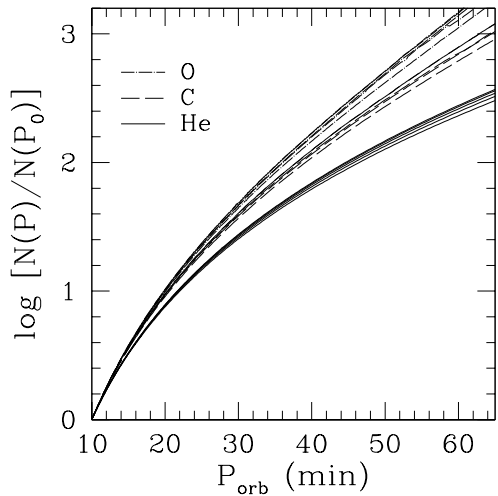}
\caption{The number distribution, $N(\Torb)$ of ultracompact systems along the adiabats shown in Figure \ref{fig:adiabmdotp}. Each distribution is normalized to 1 at $\Torb=10$ min. For each composition, from top to bottom, the tracks are for donors with initial $T_c = 10^2,\,\ee{3}{5},\,\ee{2}{6},\,\ee{5}{6}$ and $10^7$ K. The differences in the distributions between donor types are caused by the differences in their $\nad$.  The differences in $\nad$ can be seen to play a more significant role in determining $N(\Torb)$ than that of the initial donor entropy.  \label{fig:dndp}}
\end{figure}
 
While equations (\ref{eq:dlnNdlnP}) and (\ref{eq:gen_dlnNdlnP}) highlight the centrality of $\nad$ in determining $N$, to calculate $N$ for each of the adiabats in Figure \ref{fig:adiabmdotp} it is more straightforward, and slightly more accurate due to the small dependence of $N$ on $M_2/M_1$,  to calculate $\Pdot$ numerically and evaluate equation (\ref{eq:NofPdot}). We calculate $\Pdot$ from
\begin{equation}
\Pdot =- \der{\Torb}{M_2} \Mdot_2 =- \frac{P}{2 M_2} \left( 3 \nad -1\right) \Mdot_2\,,
\label{eq:pdot}
\end{equation}
where $\Mdot_2=-\Mdot$. We display the resulting $N$ distribution for each adiabatic track, normalized to the value of $N_0$ at $\Torb,_0=10$ min, in Figure \ref{fig:dndp}. It is clear that the distribution is a strong function of composition, through the differing $\nrm=\nad$, but only a weakly depends on the entropy of the donor.  The entropy dependence of $N$ derives from the fact that $\Pdot \propto M_2$ [equations (\ref{eq:RL}), (\ref{eq:mdot}), and (\ref{eq:pdot})] and donors with a higher entropy at a fixed mean density have larger $M_2$ (see Figure \ref{fig:amspRL}).  Figure \ref{fig:dndp} also shows that depending on donor type and entropy, we expect to see roughly 60-160 (for He and C/O donors, respectively) times as many systems at $\Torb \approx 40 $ min than at 10 min.  While consideration of non-adiabatic evolution will change the value of this ratio, the fact that it will be larger for C/O donors than for He donors is expected to be a robust result, regardless of the evolutionary scenario considered.  

\section{Discussion and Conclusions \label{sec:conc}}
We have presented models for arbitrarily degenerate stellar objects including Coulomb physics with masses $M<0.1 \msun$.  At these low masses, the well known \MR relations for $n=3/2$ polytropes ($R \propto M$), WDs ($R \propto M^{-1/3}$), and Coulomb dominated objects ($R \propto M^{1/3}$) merge and transition from one to another.  The connection between $T=0$ degenerate and Coulomb dominated objects has been known from the \citet{zapo69} \MR relations. The connection between polytropes and degenerate objects, neglecting Coulomb physics, is seen in the $n=3/2$ polytropes.  Our models make the final connection between the three classes of objects, filling in the gap occupied by $T\neq0$ objects in which Coulomb physics cannot be neglected.  

As discussed in \S \ref{sec:models}, a ubiquitous feature of stellar \MR relations at sufficiently high $T_c$ is the existence of a minimum mass, $\Mmin$, below which equilibrium solutions do not exist.  The cause of this is the transition from ideal gas to degenerate electrons providing the pressure support and we showed in \S \ref{sec:polytropes} that the well known maximum $T_c$ in $n=3/2$ polytropes for a fixed $M$ \citep{cox68, rapp84, stev91, burr93, usho98} is the same physics. The value of $\Mmin$ depends on both $T_c$ (in fact, $\Mmin=0$ at sufficiently low $T_c$) and the strength of Coulomb physics.  In general, the lower $T_c$ and the stronger the Coulomb interactions, the smaller $\Mmin$.  The existence of an $\Mmin$ may have a profound impact on the evolution of a donor undergoing mass loss.  For our He WDs, $\Mmin > 0$ for models with $T_c \gtrsim 10^5$ K.  As discussed in \citet{bild02}, the existence of an $\Mmin >0$ leads to the possibility of disrupting the donor through mass loss.  This could be accomplished through stable mass loss down to $M_2=\Mmin$ or via a mass transfer instability.  The latter will occur if the expansion of the donor under mass loss exceeds that of the RL.  The entropy input needed to cause this instability and the fate of the donor are the subject of future work.

We have applied our model to the accreting ultracompact MSPs.  In XTE J1751-305, the donor must be hot regardless of its composition; in XTE J0929-314 and XTE J1807-294, fully degenerate donors are possible, depending on the composition. From orbital inclination constraints, the probability that XTE J1807-294 can accommodate a  He donor is 15\%  while for XTE J0929-304, this probability is $\approx$ 35\%, providing support to the notion that some of these donors are likely C/O WDs \citep{schul01,juett01, home02}.  The evolution of each system will differ depending on the orbital inclination.  In particular, how far the system can evolve in $\Torb$ in a specified time and the expected $\Mdot$-$\Torb$ relation depends on $\sin i$ through both the mass, core temperature, and composition of the donor. In general, the $\Mdot$-$\Torb$ relation is additionally parameterized by the donor $T_c$ and composition.  We find that the number distribution of systems as a function of $\Torb$, $N(\Torb)$, is determined  by the donor's $\nrm$.  The distribution for systems with C/O donors thus varies significantly from those with He donors.  In the case of adiabatic evolution, the relative number of systems at 40 min to that at 10 min is $\approx 160$ for C/O donors and $\approx 60$ for He donors.  In addition to the accreting MSP systems, our models are applicable to the AM CVn binary systems, which are believed to be double WD binaries with a He donor \citep{warn95}.  The application of our models to these systems is the subject of current work.
 
The constant entropy models we have calculated give a lower limit on $R$ for a given $M$, $T_c$, and composition.  The actual thermal profile of a donor will depend on its prior evolution: how entropy is deposited into the star, entropy losses, and how quickly heat transport occurs as compared to mass loss.  To determine the entropy profile of a donor in a specific system requires consideration of the coupled evolution of the binary and the donor.  While our models do not address this uncertainty, they do provide limiting \MR relations based on the \emph{total} entropy content of the model in a consistent and systematic treatment without consideration of past evolution. As such they will be useful in consideration of binary systems with low-mass WD companions where time evolution of the donor's structure coupled to that of the binary itself is not computationally feasible or necessary. Tracks of our \MR relations and $\nad$ as a function of $\rho_c$ and $T_c$ for He, C, and O donors are available along with the electronic version of the paper.

We thank Deepto Chakrabarty,  Craig Markwardt, Gijs Nelemans, and Ira Wasserman for insightful and productive comments and discussions over the course of this work.  This work was supported by the National Science Foundation under grants PHY99-07949 and AST02-05956 and by NASA through grant NAG 5-8658.  L. B. is a Cottrell Scholar of the Research Corporation.

\end{document}